%Paper: cond-mat/9509127
%From: emilio@polar.fmc.uam.es (Emilio Artacho)
%Date: Wed, 20 Sep 1995 19:13:28 +0200

% ICDS 95: INTERSTITIAL OXYGEN IN SILICON AND GERMANIUM
%
\magnification=\magstep1
\hsize=17truecm
\vsize=25truecm
\hoffset=-.5truecm
\voffset=-.2truecm
\font\smrm=cmr9
\font\smbf=cmbx9
\nopagenumbers
\tolerance=5000
\parindent=20pt
\parskip=0pt

\def\PRB#1#2#3{Phys. Rev. B {\bf {#1}}, {#2} ({#3})}
\def\PRSLA#1#2#3{Proc. Roy. Soc. Lond. A {\bf {#1}}, {#2} ({#3})}

\def\SSC#1#2#3{Solid State Commun. {\bf {#1}}, {#2} ({#3})}

\centerline{\bf THEORY OF INTERSTITIAL OXYGEN IN SILICON AND GERMANIUM}

\vskip 10pt

\centerline{EMILIO ARTACHO AND FELIX YNDURAIN}

\vskip 5pt

\centerline{Departamento de F\'{\i}sica de la Materia Condensada,
C-III}
\vskip 4pt
\centerline{Universidad Aut\'onoma de Madrid, 28049 Madrid, Spain.}

\vskip 15pt

\noindent
{\smbf Keywords:} {\smrm Oxygen in covalent semiconductors.}

\vskip 20pt

\noindent
{\bf Abstract.}
The interstitial oxygen centers in silicon and germanium are reconsidered
and compared in an analysis based on the first-principles total-energy
determination of the potential-energy surface of the centers, and a
calculation of their respective low energy excitations and infrared
absorption spectra. The total-energy calculations reveal unambiguously that
interstitial oxygen is quantum delocalized, the delocalization being
essentially different in silicon and in germanium. Oxygen
in silicon lies at the bond center site in a highly anharmonic
potential well, whereas in germanium it is found to rotate almost freely
around the original Ge-Ge bond it breaks. This different delocalization is
the origin of the important differences in the low energy excitation spectra:
there is a clear decoupling in rotation and vibration excitations in germanium,
giving different energy scales (1 cm$^{-1}$ for the rotation, 200 cm$^{-1}$
for the $\nu_2$ mode), whereas both motions are non-trivially mixed in silicon,
in a common energy scale of around 30 cm$^{-1}$. The calculation of the
vibrational spectra of the defect reveals the existence of vibrational modes
(related to the $\nu_1$ mode) never been experimentally observed
due to their weak infrared activity. It is found that the combination of
these modes with the well established $\nu_3$ asymmetric stretching ones
is the origin of the experimentally well characterized modes at frequencies
above the $\nu_3$ mode frequency.

\vskip 20pt

\noindent
{\bf I. Introduction}

\vskip 8pt

\noindent
It is broadly accepted that interstitial oxygen in silicon (Si:O$_i$) [1] and
in germanium (Ge:O$_i$) [2-5] represent very similar centers in
their geometry and dynamics. The aim of this paper is to show the contrary:
the quantitative differences in the low-energy experimental results for
Si:O$_i$ [6] and Ge:O$_i$ [3] are shown to correspond to fundamentally
different realities, that, in turn, lead us to predict substantial differences
in their infrared absorption spectra. The basic physics of both centers as
well as their similarities and differences are analyzed and discussed in this
work, the essence being graphically depicted in Fig.~1.
Oxygen~breaks~a~covalent~bond

\vskip 1.5truecm \noindent \hskip 10truecm \vbox{ \hsize 7truecm \noindent
\smbf Fig. 1: \smrm Schematic representation of the oxygen delocalization
in Si:O$_i$ and Ge:O$_i$, in the plane perpendicular to the Si-Si (Ge-Ge)
axis.} \vskip 1.0truecm

\noindent
between two semiconductor atoms and remains quantum delocalized between them.
The delocalization is essentially different in Si:O$_i$ and in Ge:O$_i$, both
being again different from the trivial harmonic motion around a well defined
geometry. In the case of silicon [7,8], oxygen is delocalized in the close
neighborhood of the original silicon-silicon bond center (BC) position, whereas
in germanium the defect can be very precisely accounted for by an almost free
rotor. As a consequence of the different geometries the low energy excitation
spectra are quite different. In addition, in the infrared spectral region,
interstitial oxygen in silicon induces a backbond-stretching resonance at
517 cm$^{-1}$ which has no counterpart in Ge:O$_i$, while in germanium there
is a $\nu_2$ bending mode in the infrared that in Si:O$_i$ appears among the
low-energy excitations.

In this work we address at these differences between Si:O$_i$ and
Ge:O$_i$ from different and complementary points of view.
With the help of total energy calculations, we determine the "equilibrium"
configuration of Si:O$_i$ and Ge:O$_i$ and analyze the low energy
excitations experiments. The atomic vibration and infrared spectra
associated to the defect are also calculated. We find vibrational modes
which have never been directly observed experimentally. We show that
these modes, in combination with the high frequency stretching
modes, are responsible for the experimentally observed modes induced
by oxygen at 1751 $cm^{-1}$ and 1260 $cm^{-1}$ in Si:O$_i$ [9] and
Ge:O$_i$ [2], respectively.

\vskip 20pt

\noindent
{\bf II. Atomic configuration and far infrared analysis}

\vskip 8pt

\noindent
In Fig.~2 the results of the analysis of the low energy excitations of the
interstitial centers are shown. In the case of Si:O$_i$ the theoretical
results of Yamada-Kaneta {\it et al.} [7] along with the experimental
results of the far infrared measurements of Bosomworth {\it et al.} [6]
are displayed, supporting the image of Fig.~1. In the case of Ge:O$_i$,
the experimental results of Gienger {\it et al.} [3] are drawn along with the
results of our hindered rotor theoretical analysis, where the Hamiltonian of
an elastic rotor [10] ($H_{\rm o}=B l^2 - D l^4$, $l$ being the angular
momentum) is perturbed by an angular-dependent potential $H^\prime=(A/2)
\cos(6\phi)$ with an amplitude $A\approx 0.6$ meV. The model is solved exactly
to the required precision, giving a remarkable agreement with experiments,
which supports the picture of Fig~1. It has to be stressed, however, that
higher harmonics of the angular potential affect very little the displayed
results, leaving the hindering potential (and thus the angular energy barrier,
and the position and number of minima) indetermined in that respect (the
measurement of the splitting of the $l=\pm 6$ levels would resolve the
indetermination). This is illustrated in Fig.~2, where an added $\cos(12\phi)$
term is considered. Note the possible qualitative differences.

\vskip 1.5truecm \noindent \hskip 10truecm \vbox{ \hsize 7truecm \noindent
\smbf Fig. 2: \smrm Low-energy excitations of Si:O$_i$ and Ge:O$_i$. For
Si:O$_i$ far-infrared-absorption data [6] are compared with the fit of
Yamada-Kaneta {\it et al.} [7] based on the anharmonic potential well coupled
to the $\nu_3$ vibration mode. Also displayed is the renormalized potential
for the ground state of $\nu_3$. For Ge:O$_i$, the results of phonon
spectroscopy of Gienger {\it et al.} [3] are compared with the results of
the hindered rotor model. Also shown are three potentials compatible with
the data, differing in the amplitude of the $\cos(12\phi)$ term (zero,
negative, or positive, respectively).} \vskip 1.2truecm

   It is important to note in Fig.~2 the different energy scales involved.
The angular delocalization of oxygen in germanium gives rise to low lying
rotation excitations which are essentially decoupled from the vibrational modes
at a different energy scale (see below). In silicon, radial and angular
excitations are non-trivially mixed in a common energy scale.

   First-principles total-energy calculations have been performed in the
manner described elsewhere [8], using a cluster-Hartree-Fock approximation.
The results of the calculations, carefully considering relaxations up to
second nearest neighbor atoms of oxygen, are shown in Fig.~3 for the radial
dependence. The size of the barrier for the radial motion in Ge:O$_i$
localizes the oxygen distance to BC around $r_{\rm O}=0.58$ \AA. The amplitude
of the $\cos(6\phi)$ harmonic for the angular dependence of the potential has
been calculated to be of a few tenths of a meV for Ge:O$_i$, in qualitative
agreement with the model described above.These results give additional support
to the picture of Fig.~1. In the equilibrium geometry the Ge-O-Ge (Si-O-Si)
angle is 140$^{\rm o}$ (180$^{\rm o}$) and the Ge-O (Si-O) distance
1.70\AA\ (1.56\AA).

\vskip .7truecm \noindent \hskip 10.5truecm \vbox{ \hsize 6.5truecm \noindent
\smbf Fig. 3: \smrm Radial potential for the oxygen motion for Si:O$_i$ and
Ge:O$_i$ obtained from first principles relaxing up to second nearest
neighbors of oxygen for every point. There is a flat region in the Si:O$_i$
case [8] compatible with the potential in Fig.~2.} \vskip 1.2truecm

\vskip 20pt

\noindent
{\bf III. Vibrational modes and infrared absorption}

\vskip 8pt

\noindent
The infrared absorption spectrum of Ge:O$_i$ is calculated for a fixed polar
angle, in absolute units [11]
and the results are shown in Fig.~4 (the results for Si:O$_i$ have been
published elsewhere [8]). Four distinct features appear at 877, 416,
230, and 0 cm$^{-1}$, respectively. The latter corresponds to the free
rotation of oxygen and the other three correspond to the $\nu_3$ stretching,
$\nu_1$ bending, and $\nu_2$ rocking modes of the Ge$_{2}$O unit [1] (see
below).~The~never~ob-

\vskip 2.5truecm \noindent \hskip 10truecm \vbox{ \hsize 7truecm \noindent
\smbf Fig. 4: \smrm Infrared absorption coefficient calculated for Ge:O$_i$
for an oxygen concentration [O]$=10^{18}$ cm$^{-3}$. The inset shows a
magnification of the spectral region corresponding to the upper Ge phonon
bands, compared with the equivalent in Si:O$_i$.} \vskip 2.5truecm

\noindent
served $\nu_2$ peak appears as a resonance in the germanium
continuum and corresponds to the radial vibration of the rotor. Indeed, the
centrifugal distortion coefficient $D=4 B^3/ \hbar^2 \omega^2$ [10] in the
hindered rotor model gives a value for $\omega$ of the magnitude of the
frequency of this $\nu_2$ mode (the exact value depends on the amplitude of
the $\cos(12\phi)$ term). The position of this $\nu_2$ feature is not as
well defined by our model as the others, being more sensitive to small
changes in the dynamical matrix.

\vskip 6.0truecm \noindent \hskip 10.8truecm \vbox{ \hsize 6.2truecm \noindent
\smbf Fig. 5: \smrm Displacement-displacement correlation functions,
$\langle u_{i}u_{j} \rangle (\omega)$ of oxygen and neighboring germanium
atoms in Ge:O$_i$.} \vskip 2.2truecm

Although the asymmetric streching mode at 877 cm$^{-1}$ has been
very well characterized experimentally as a main absorption peak at 860
cm$^{-1}$, the mode at 416 cm$^{-1}$ (parallel to the 596 cm$^{-1}$ mode in
Si:O$_i$) has never been observed experimentally. This is due to the weakness
of this feature and also to the fact that the infrared spectrum of
crystalline germanium shows a two-phonon infrared absorption in this frequency
range [12] that can hide the absorption of O$_i$ at
the usual concentration. Nevertheless, the combination of
this mode with the asymmetric stretching at 877 cm$^{-1}$ has been
observed experimentally at 1260 cm$^{-1}$ [2], in the same manner as
in Si:O$_i$ the non-infrared active mode at 596 cm$^{-1}$ combines
with the asymmetric stretching one to give the 1751 cm$^{-1}$ band [9].
This assignment of the combination mode is soundly supported by
the agreement of the isotope shifts predicted by the theory and the
corresponding experimental results. These data have already been presented
for Si:O$_i$ [8,13] and will be published elsewhere for Ge:O$_i$ [14].

   The displacement-displacement correlation functions
$\langle u_{i}u_{j} \rangle (\omega)$ of atoms in the vicinity of the defect
are shown in Fig.~5, confirming the assignments made above.
%Also the mean square displacement of the atoms at the different
%vibrational modes are given in Table I. (****)
The inset of Fig.~4 shows the infrared absorption
near the bulk continuum for both Si:O$_i$ and Ge:O$_i$. The figure
shows remarkable differences between these two spectra.
The asymmetric resonance in Si:O$_i$ at 517 cm$^{-1}$, absent in Ge:O$_i$,
is due to the backbonding relaxation caused by the incorporation of oxygen.
This relaxation is larger in silicon than in germanium.
For a better understanding of this feature we have calculated the
displacement-displacement correlation function between the silicon
atoms forming the backbonds. Results of the calculations are shown
in Fig.6. These results show very explicitly the origin of the
modes at 517 and 596 cm$^{-1}$ (416 cm$^{-1}$ in Fig.~5) in silicon
(germanium). The mode at 517 cm$^{-1}$ is a silicon optic mode (transverse
to the defect axis), with a higher frequency due to the backbonding
compression. The mode at 596 cm$^{-1}$ (416 cm$^{-1}$) is a
(longitudinal) optic mode split-off the band due to the presence
of the stronger Si-O-Si (Ge-O-Ge) bond in the semiconductor
lattice.

\vskip .8truecm \noindent \hskip 6truecm \vbox{ \hsize 11truecm
\vbox{ \hsize 9truecm \noindent
\smbf Fig. 6: \smrm Displacement-displacement correlation~func- tions
of silicon atoms forming backbonds in Si:O$_i$.}
%} \vskip 1.2truecm

\rm

\vskip 20pt

\noindent
{\bf IV. Concluding remarks}

\vskip 8pt

\noindent
Total energy calculation, analysis of the low energy excitations
and calculation of the infrared spectra have allowed us to fully
understand the similarities and differences of interstitial oxygen
in silicon and germanium. It is found that, in spite of the
similarity of the defects, there are fundamental differences that
manifest themselves in the low energy excitation spectra and appear
as subtle but relevant details in the infrared absorption.

\vskip 8pt

We thank A. J. Mayur for bringing our attention to the problem and B. Pajot
for fruitful discussions. This work has been supported in part by the
DGICYT through grant PB92-0169.}

\vskip 10pt

\noindent
{\bf References}

\vskip 8pt

\noindent

\item{[1]}See R.C.Newman, {\it Infrared Studies of
Crystal Defects,} Taylor~and~Francis,~London,~1973.
\item{[2]}B. Pajot and P. Clauws, {\it Proc. of the 18$^{th}$
International Conference on the Physics of Semiconductors}, ed. O. Engstr\"om,
(World Scientific, Singapore, 1987) p. 911.
\item{[3]}M. Gienger, M. Glaser and K. La{\ss}mann, \SSC{86}{285}{1993}.
\item{[4]}A. Liz\'on-Nordstr\"om and F. Yndur\'ain, \SSC{89}{819}{1994}.
\item{[5]}A. J. Mayur, M. D. Sciacca, M. K. Udo, A. K.
Ramdas, K. Itoh, J. Wolk, and E. E. Haller, \PRB{49}{16293}{1994}.
\item{[6]} D. R. Bosomworth, W. Hayes, A. R. L. Spray, and G. D. Watkins,
\PRSLA{317}{133}{1970}.
\item{[7]}H. Yamada-Kaneta, C. Kaneta, and T. Ogawa,
\PRB{42}{9650}{1990}.
\item{[8]}E. Artacho, A. Liz\'on-Nordstr\"om and F. Yndur\'ain,
\PRB{51}{7862}{1995}; E. Artacho and F. Yndur\'ain, {\it Proc. of the
22$^{nd}$ International Conference on the Physics of Semiconductors}, ed.
D.~J.~Lockwood, (World Scientific, Singapore, 1995) p. 2459.
\item{[9]}B. Pajot, H.J. Stein, B. Cales and C. Naud, J. Electrochem.
Soc. {\bf 132}, 3034 (1985).
\item{[10]} G. Herzberg, {\it Infrared and Raman Spectra of Diatomic
Molecules,} (D. Van Nostrand Company, Inc., 1945) p. 104.
\item{[11]}Model dynamical-matrix calculation for a single O impurity in an
infinite semiconductor, described in Ref. 8. The absolute units are obtained
as in E. Mart\'{\i}nez and M. Cardona, \PRB{28}{880}{1983}.
\item{[12]}R. Tubino, L. Piseri, and G. Zerbi, J. Chem. Phys. {\bf 56}, 1022
(1972); K. Winer and M. Cardona, \PRB{35}{8189}{1987} and references therein.
\item{[13]}B. Pajot and P. Cales, Mat. Res. Symp. Proc. {\bf 59}, 39 (1986).
B. Pajot, E. Artacho, C. A. J. Ammerlaan, and J. M. Spaeth, to appear in
J. Phys.: Condens. Matter.
\item{[14]}E. Artacho, F. Yndur\'ain, B. Pajot, L. I. Khirunenko,
R. Ram\'{\i}rez, and C. P. Herrero, to be published.

\noindent

\vfill
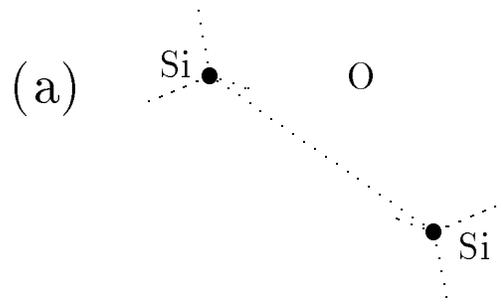
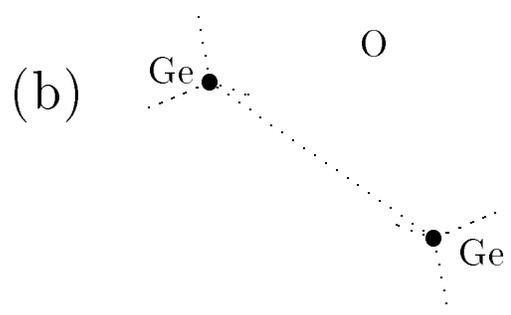
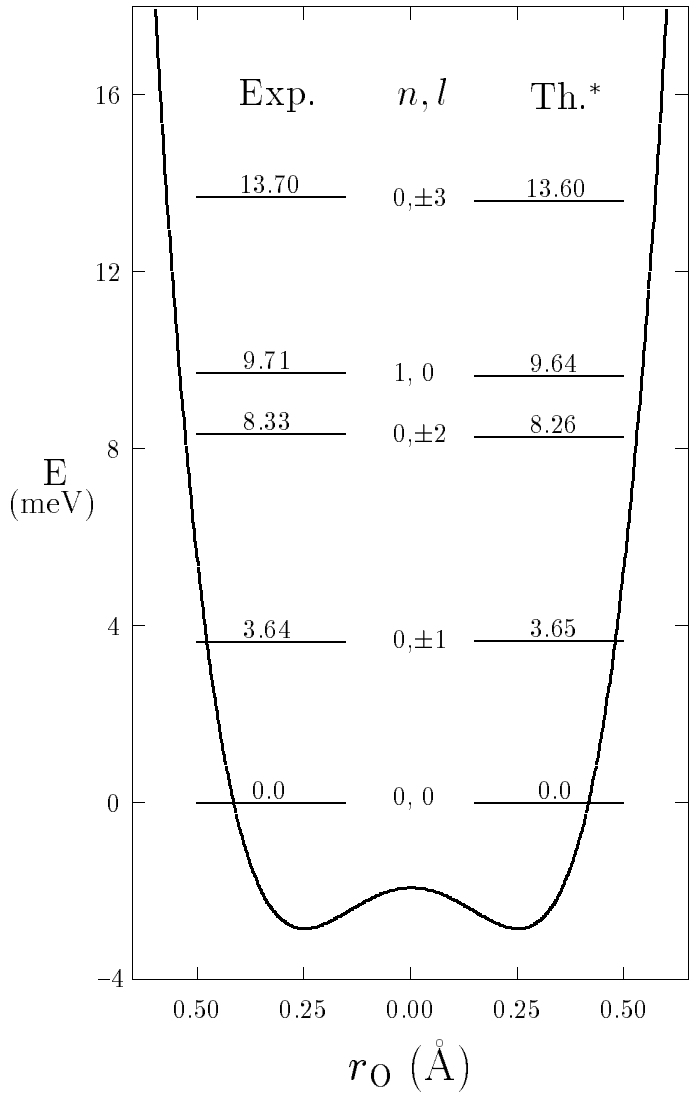
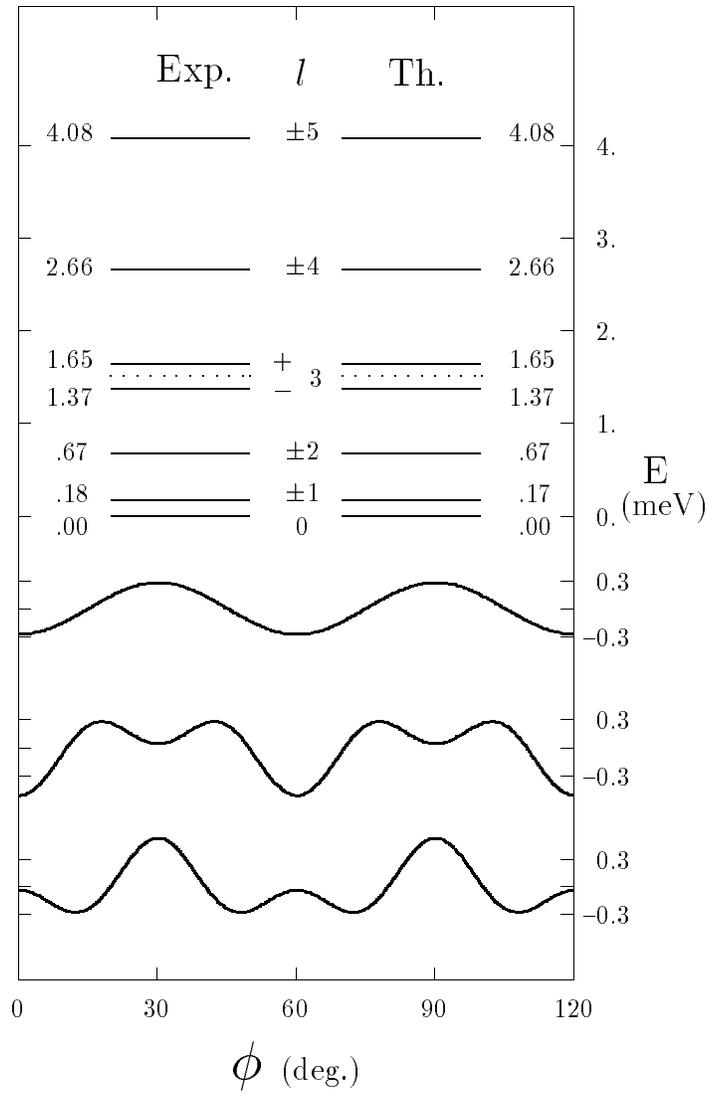
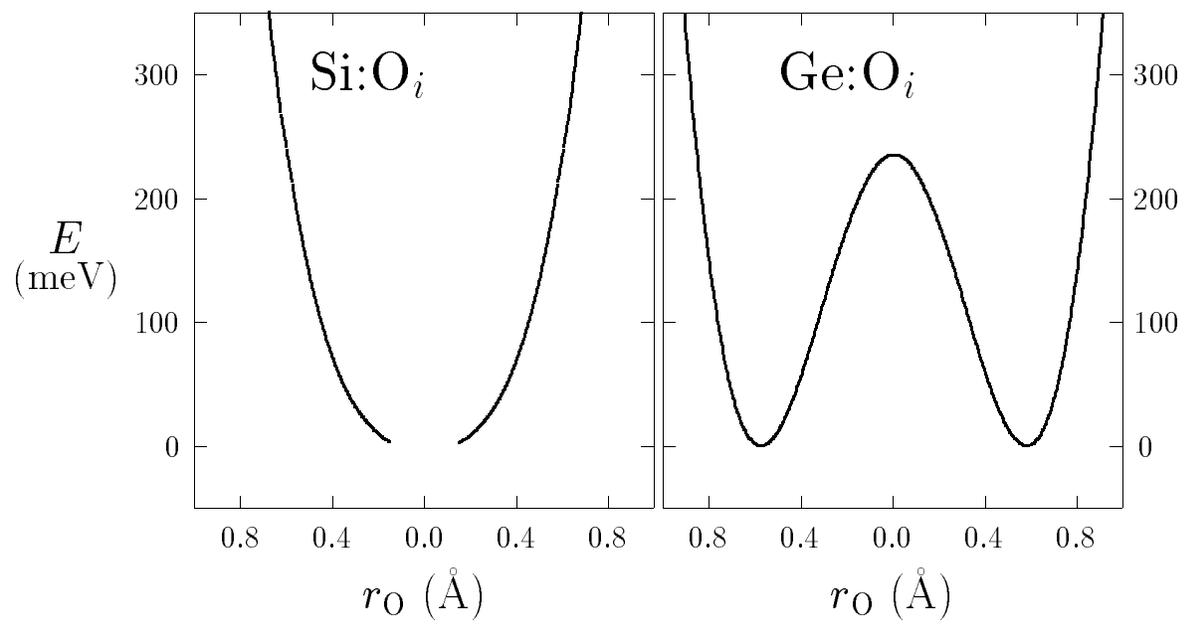
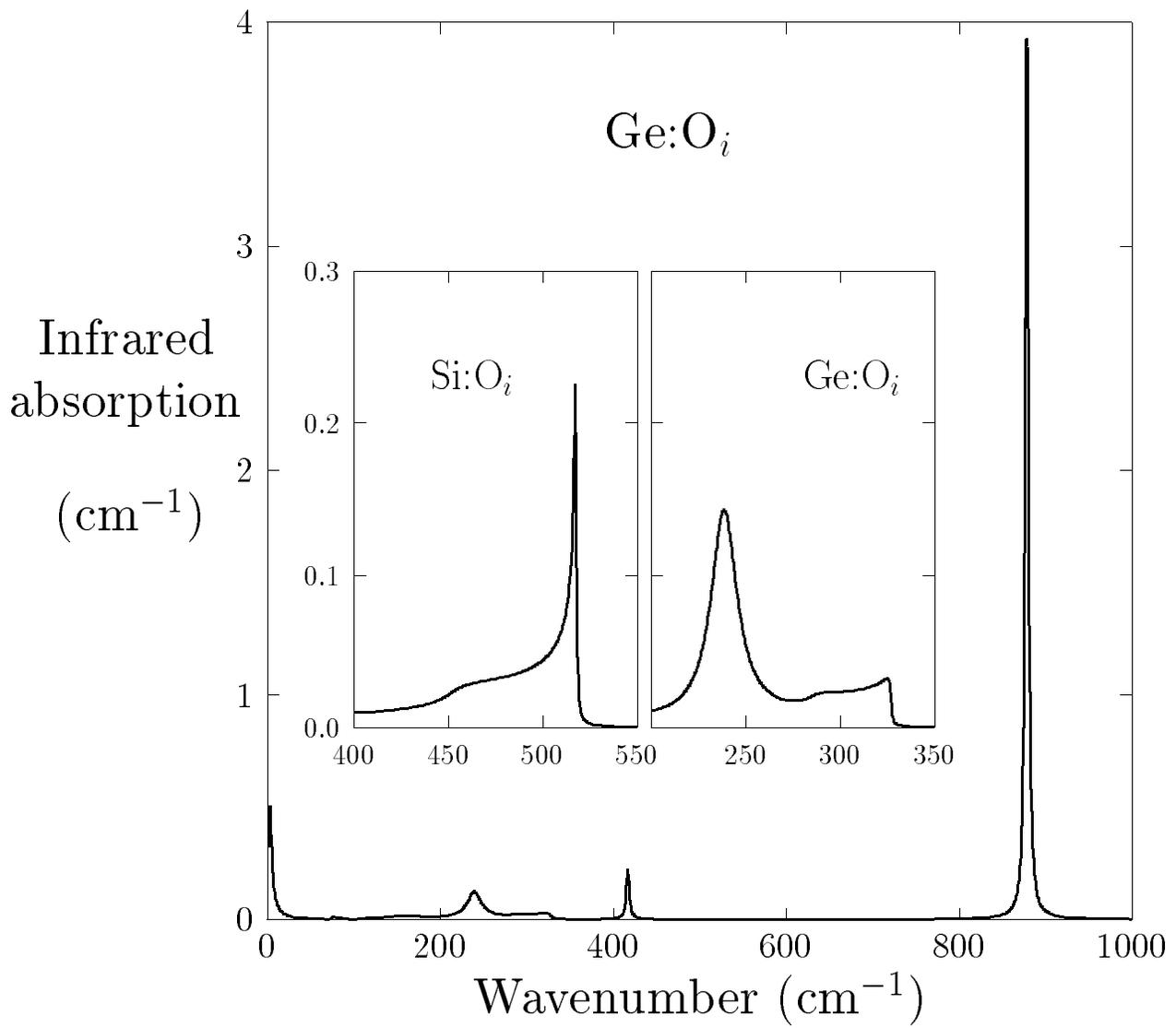
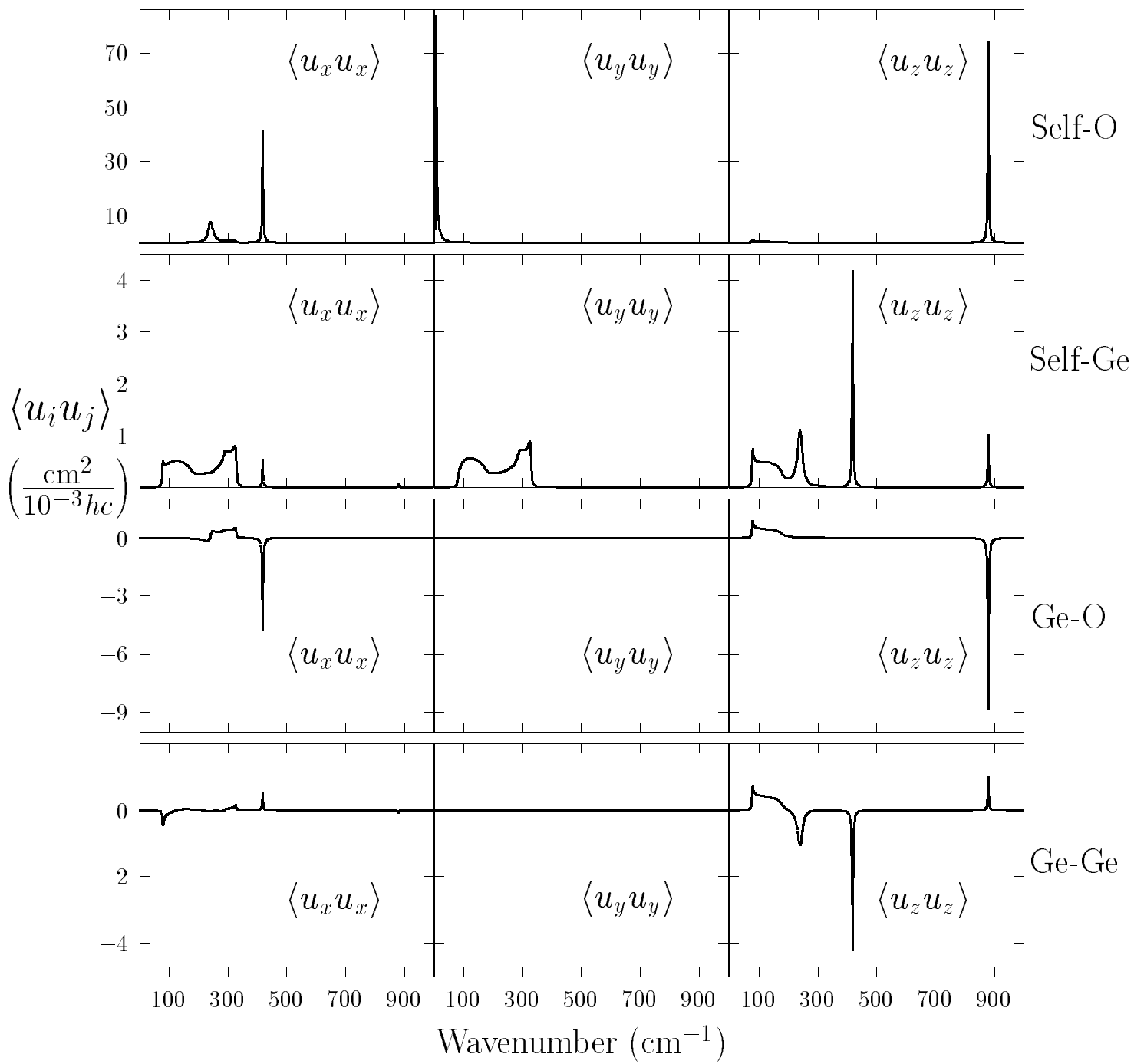
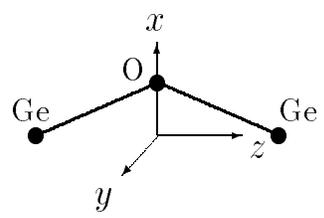
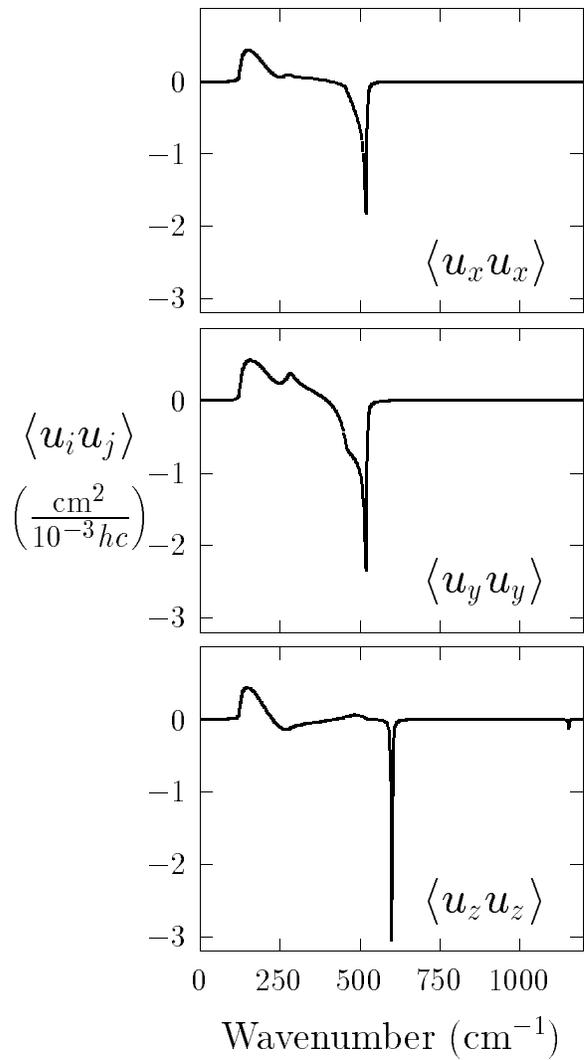
\end